\documentclass[aps,prl,twocolumn,showpacs,]{revtex4}

\usepackage{epsfig}

\begin{document}

\title{Inverse slow relaxation in granular hopping systems}
\author{N. Kurzweil}
\author{A. Frydman}

\address{The Department of Physics, Bar Ilan University, Ramat Gan 52900, Israel}

\begin{abstract}
We present experimental results that demonstrate a glassy behavior
in the conductance of quench condensed insulating granular metals
that is different from that observed in continuous disordered
systems. Exciting the granular system by biasing the sample with a
high electric field results in a slow conductance change both during
the excitation and during the relaxation back to its steady state.
The time scales for these processes are many orders of magnitudes longer than the typical hopping time. We find that, initially, this conductance change has an opposite sign to
that observed in similar experiments performed on continuous films.
Only after relatively long times this trend is reversed and the
samples exhibit conventional behavior. We suggest that the granular
systems exhibit a new glassy process related to charge
redistribution among the grains. This process combines with the
relaxation processes that are characteristic of disordered systems
to give rise to a unique relaxation profile.

\end{abstract}

\pacs{73.23.-b; 73.40.Rw; 75.50.Cc}

\date{\today}
\maketitle

Glassy behavior of the conductivity of strongly disordered systems
has attracted attention for over a decade. Exciting a hopping system
out of its thermal equilibrium leads to an increase in conductivity.
The relaxation processes toward equilibrium are characterized by
extremely long times, memory phenomena and aging effects. The
majority of experiments were performed on Indium Oxide thin films
which were excited by applying a gate voltage, thus abruptly
changing the carrier concentration \cite{moshe,moshe1,ady1,ady1_5,ady2}.
Following the sharp initial increase, the
conductivity exhibited a logarithmic  time decay to a new
equilibrium state with time scales of the order of many hours or
days. This slow conductivity relaxation showed many phenomena
characteristic of glasses such as scaling behavior in aging
experiments and memory effects.

Very similar glassy behavior (e.g., slow relaxation, memory and
aging) was reported also in granular Au or Al \cite{adkins, grenet}
demonstrating that this phenomena is not unique to Anderson
insulators. This is not surprising since, in principle, the
electrical transport in  a granular metal in the insulating phase is
very similar to that of a disordered continuous metal. The main
difference is that in a granular system the electrons are localized
on isolated grains instead of localized electronic states created by
the random potential. Each grain is a small metal and can contain a
large number of electrons, unlike the single localized state.

Recently, Orlyanchik and Ovadyahu \cite{vladic} used a different
technique to excite Indium oxide samples in the hopping regime. They
biased the sample by applying a large electric field, thus driving
the system out of its ohmic regime, and studied the time-dependent
conductance both during excitation (while applying the voltage) and
relaxation (after the field was removed). They found that the
application of the field caused a fast conductance increase followed
by a slow rise towards an asymptotic value. Removing the field
caused a fast conductance decrease followed by a very slow drop
toward the thermal equilibrium.

In this letter we report on an experiment in which we apply a
similar technique to that of Orlyanchik and Ovadyahu  to excite a
granular metal. Surprisingly (unlike the case of applying a gate
voltage), we find that applying a bias voltage gives rise to an
opposite trend to that seen in the continuous disordered systems. We
attribute this difference to the unique electric field distribution
in granular metals.

The samples studied in this work were various (Ag, Au and Ni) thin
granular films fabricated using the quench condensation technique.
This method enables growing and measuring a 2D granular sample at
low temperature and high vacuum without exposing it to the
atmosphere or thermally cycling it \cite{qc1, qc2, qc3}. Prior to
the cool-down, two Gold leads were evaporated on a insulating Si/SiO
or glass substrate. The substrate was then placed in a UHV system
and immersed in liquid He. An ultrathin metal film was then
evaporated on the cold substrate, forming a discontinuous, granular
structure with grain sizes of 100-200 {\AA} and intergrain distances
of a few tens of {\AA}. Sample resistance was monitored during the
growth and the process was stopped at a desired resistance. We
studied samples with resistances in the range of $50$ $M\Omega-5$
 $G\Omega.$ After terminating the film deposition the sample was
allowed to relax for a few hours until the conductance change was
smaller than 0.1 $\%$/hour and a base conductance line was
determined. Then, a large electric field, $F_{n} =2-600$ $V/cm$, was
applied abruptly and the time-dependent-conductance, G(t), was
monitored. After a desired waiting time, $t_{w}$, the field was
abruptly reduced to $F_{0} \approx 0.5$ $V/cm$ and the low voltage
relaxation G(t) was measured. All conductance measurements were
performed by measuring the current through a small series resistor
(at least two orders of magnitude smaller than the sample
resistance), similar to the method described in ref \cite{vladic}.

Typical excitation and relaxation G(t) profiles for two different
samples are shown in fig. 1.  It is seen that the initial trend was
opposite to that observed by Orlyanchik and Ovadyahu \cite{vladic}.
During the excitation the conductivity decreased with time instead
of the increase observed in continuous films, and during the
relaxation process the conductivity increased slowly contrary to
decrease observed in continuous films. After a period of time that
depended on $t_{w}$ (typically, a few hundred to a few thousand
seconds), the trend reversed and the conventional behavior typical
of Anderson insulators was observed. This behavior was observed in all investigated granular metals without significant dependence on the material.

\begin{figure}
{\epsfxsize=2.5 in \epsffile{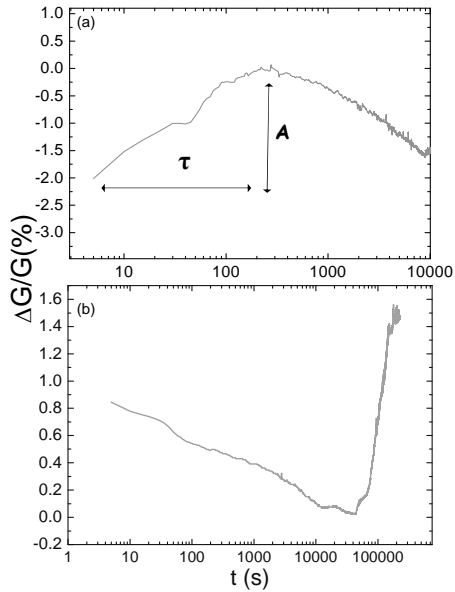}} \caption{ Relative
conductance change versus time for two Ag granular films excited
with a bias field $F_{n}$=200 V/cm. (a) relaxation curve under
$F_{0}$=0.5 V/cm, R=360 M$\Omega$, T=4.2 K. (b) the relative
conductance during the excitation, R=80 M$\Omega$.  \small}
\label{particle}
\end{figure}

Since the behavior during excitation and relaxation is similar (with
opposite signs) for the reminder of this paper we will focus on the
processes of relaxation after switching off the electric field. In
figure 1 we plot the change in conductivity versus time, $\Delta
G(t)/G=(G(t)-G_{0})/G_{0}$, where $G_{0}$ is the extremum
conductance at which the change in the trend occurred and t=0 when
the excitation was introduced or removed. We define two parameters,
(as seen in fig.1a): $\tau$, the time it takes the conductivity to
reach $G=G_{0}$ and A, the change in conductivity during $\tau$. In
all measurements we identify two regimes of the relaxation G(t)
profiles: the initial conductance increase at $t < \tau$ we name the
"opposite effect" and the conductance decrease at $t > \tau$ which
is the conventional behavior similar to the results in continuous
insulators. We note that a typical sample had a resistance of $\sim
100$ $M\Omega$ and the measured circuit capacitance was $\sim$ $100$
$pF$ yielding $RC \sim 10$ $ms$, which is many orders of magnitude
smaller than the typical measured relaxation time $\tau$.

\begin{figure}
{\epsfxsize=2.6 in \epsffile{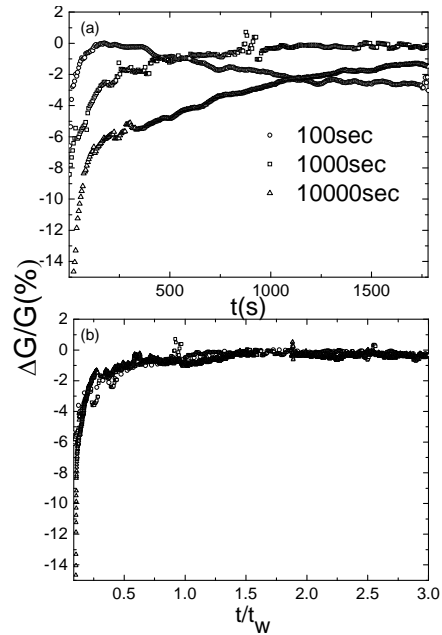}} \caption{(a) Conductance
during relaxation for different values of $t_{w}$ (b) The same data
as in (a) plotted as a function of t/$t_{w}$.  R=5.4 G$\Omega$,
$F_{n}$=100 V/cm. The conductance oscillations which appear in the
graphs are rather reproducible and probably result from the
mesoscopic nature of the sample \cite{future}.\small}
\label{particle}
\end{figure}

In order to study the glassy nature of the "opposite effect" we
performed aging experiments using the following scheme: We
stabilized the system at low applied electric field, $F_{0}$, for
several hours, thus reaching a steady state of the system. Then we
excited the system by applying high bias field, $F_{n}$, for a
waiting time, $t_{w}$ after which the field was reduced back to
$F_{0}$. Conductivity measurements were performed for a time of the
order of $3\times t_{w}$ after removing the high field. The
relaxation curves of the conductivity versus time for three
different $t_{w}$ are plotted in fig. 2(a). The same data is plotted
in fig. 2(b) as a function of $t/t_{w}$. It is seen that all curves
collapse onto a single curve as a function of $t/t_{w}$ giving rise
to simple scaling behavior \cite{ady1_5,rem1}.

\begin{figure}
{\epsfxsize=2.6  in \epsffile{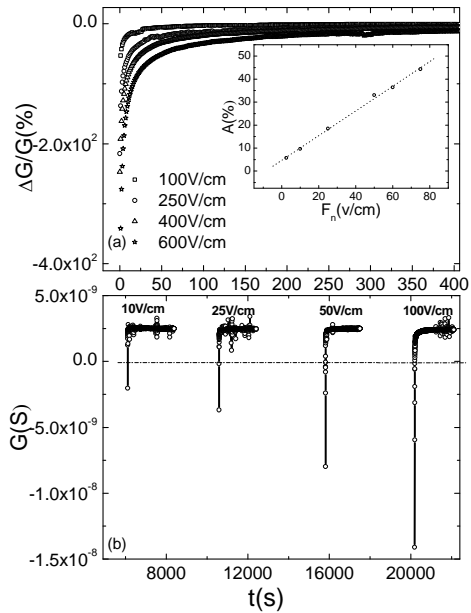}} \vspace{-1 cm}\caption{(a)
Relaxation curves after excitations by different $F_{n}$.
$t_{w}$=2000 sec and R=150 M$\Omega$. The inset shows the amplitude,
A, as a function of applied bias voltage for a 320 M$\Omega$ sample
. (b) Relaxation processes where "negative conductance" was
obtained, R=250 M$\Omega$.\small} \label{particle}
\end{figure}

Varying the strength of the exciting field also had a significant
effect on the G(t) profiles.  Fig. 3a which depicts $\Delta G/G$ for
different $F_{n}s$ shows that the relative conductance change, A,
grows linearly  with $F_{n}$. For high enough $F_{n}$, switching off
the field gave rise to a short period of "negative conductance" (see
fig. 3b), indicating that, for short times, the current flow was
opposite to the voltage direction. We found that $\tau$ does not
depend on the value of the exciting field and no correlation was
found between A and $t_{w}$.

All the above results were obtained employing a DC voltage
excitation. We performed similar measurements using an AC voltage
excitation and measured the AC current through the sample. Fig.4
shows that exciting the samples with an AC voltage, even with
frequencies as low as 2 Hz, caused the "opposite effect" to
disappear, leaving only the conventional, Orlyanchik and Ovadyahu
relaxation trend.

\begin{figure}
{\epsfxsize=2.5 in \epsffile{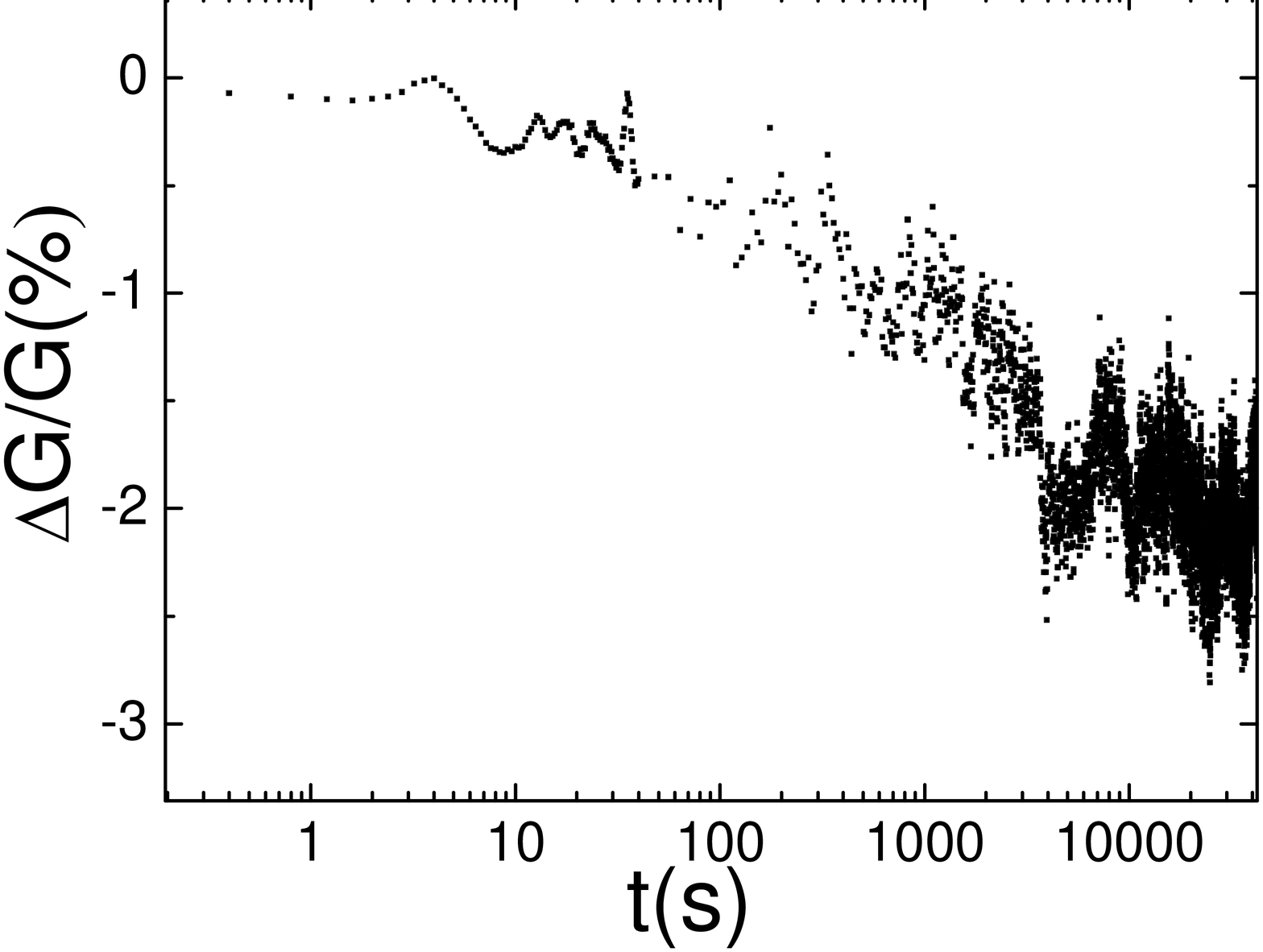}} \caption{The excess
conductance versus time after switching off a 2 Hz AC field of 25
V/cm.\small} \label{particle}
\end{figure}
In the following paragraphs we propose a model for the observed
results, based on the granular nature of the samples. The
conductivity in these systems is governed by electron hopping
between the metal grains \cite{aviadold}. When an external electric
field $F_{n}$ is applied to the system the charge on each grain
reorganizes itself in order to reach electrostatic equilibrium. If
the metallic grains are far from each other the charge on the grains
is fixed and is redistributed on the grain so that the
internal field cancels out the external $F_{n}$. This process occurs
over a characteristic
metallic screening time (typically less than 1 ps). However,
for closely packed grains, strong electrons interactions cause the
rearrangement of the charge distribution in each grain to affect the
arrangement of its neighbors. Therefore both the magnitude and the
direction of the screening internal field in each grain change in
time, mostly by hopping events, until it reaches a steady state.
This process is interactive - hierarchic, hence is characterized by
very long times leading to a slow decrease of the electric current
in the system. After the excitation is removed, the grains discharge
in order to adapt to the new electrostatic potential. This too is a
slow process leading to a long-time current decrease, in which the
direction of the current is opposite to that of the original
excitation. This leads to a new glassy process which competes with
the conventional relaxation processes of a hopping system relaxing
to equilibrium.

In order to test this model we simulated the time dependence
conductivity in a one dimension granular system. We
consider a chain of metallic grains connected to two large leads
which are biased by a voltage, V. Each grain is connected to its
neighbor by junctions of capacitance $c_{i}$ and tunneling
resistance $R_{i}$, as illustrated in the inset of fig 5. The
potential, $\varphi_{i}$, of each grain is given by

\begin{equation}
    \varphi_{i}=\frac{\sum c_{j}\varphi_{j}+n_{i}e}{c_{\Sigma}},
\end{equation}

where j is the index of the neighboring capacitors and grains,
$n_{i}$ is the number of the electrons added to the grain i from the
initially neutral condition and $c_{\Sigma}$ is given by $\sum
c_{j}$. The potentials of the two leads are 0 and V. The total
electrostatic energy of the system is given by:
\begin{equation}
U=\sum U_{i}=\frac{1}{2}\sum c_{i}(\varphi_{i+1}-\varphi_{i})^{2}.
\end{equation}

By subtracting the total coulomb energy of two different
combinations of $n_{i}$ and V (state 1 and state 2), we find $\Delta
E=U_{2}-U_{1}$, the total coulomb energy cost of the system
transferring from one state to the other. However, the charge
distribution in the system changes in order to keep a constant bias
voltage, therefore we add to $\Delta E$ the energy supplied by the
voltage source when an electron tunnels through any capacitor.

We define a vector $\overrightarrow{\rho} (V, n_{1},....,n_{i},t)$
which is the probability to have $n_{1}....n_{i}$ electrons on
grains 1....i respectively, under bias voltage V at time t.

 At non zero temperature, there is a probability of tunneling processes, for
which the rate of change from one distribution to another is:
\begin{equation}
\Gamma({n_{1}^{1}...n_{i}^{1}},{n_{1}^{2}...n_{i}^{2}})=\frac{1}{Re^{2}}(\frac{-\Delta
E}{1-exp(\Delta E/k_{B}T)}),
\end{equation}
where $n_{1}^{k}...n_{i}^{k}$ is the charge distribution in the
state k and R is the resistance of the junction.

The master equation for the system is thus given by:

\begin{equation}
\frac{\partial\overrightarrow{\rho}(V,n_{1},....,n_{i},t)}{\partial
t}=
\Gamma\cdot\overrightarrow{\rho}(V,n_{1},....,n_{i},t),\end{equation}

where $\Gamma$ is the rate matrix and $\overrightarrow{\rho}$ is the
state vector. Because the system is not in equilibrium, the current
through each junction is not the same. The measured current is the
sum of all the possibilities to tunnel through the last junction.
Our initial conditions are probability 1 for zero electrons on each
grain. We then introduce a high bias voltage V for a time of
$t_{w}$. We solve eq. 4 and calculate the current for every step.
Fig. 5 shows the relaxation curve of the current after switching off
the high voltage excitation so that V=0. The current is negative and
has an amplitude which decreases to zero over a time which is much
longer ($\sim$3 orders of magnitude) than a typical hopping process
even for a system of 5 grains. The inset of fig 5 shows that the
characteristic relaxation times increase monotonically as a function
of the number of grains, hence the times can be expected to be much
longer in a macroscopic sample. We note that  this is clearly a
simplistic model which takes into account only nearest-neighbor hops
and doesn't include the disorder properties of the system but we
believe that it conveys the essential physics of the
 slow, glassy dynamics responsible for the time dependent
conductivity.

\begin{figure}
{\epsfxsize=2.6 in \epsffile{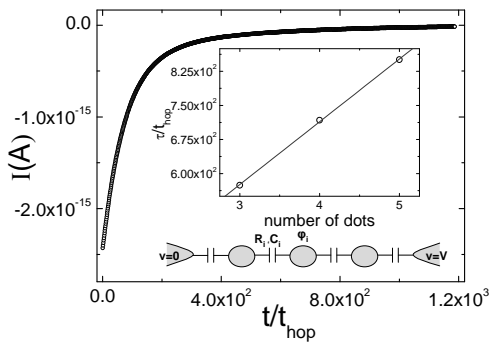}} \caption{Simulation of the
current versus time, normalized to the hopping time, after an
excitation with bias voltage V=500 V of a 5 grains chain.
$C_{i}=10^{-17}$ $F, R_{i}=10^{8}$ $M\Omega$. The inserts show a
schematic description of the model and the dependence of the
characteristic relaxation time on the number of dots in the chain.}
\label{particle}
\end{figure}

In the actual experiment the electric field is not reduced to zero
but rather to a small dc field for conductivity measurements. So the
negative current observed in the simulations has to be added to a
finite positive current. The sum of those two can be expected to
lead to a current profile similar to those seen in figure 3b in
which the total conductivity may begin negative but eventually approaches a steady state value of
$G_{0}$. The change in the conductance is obtained as a response to the DC applied field. When the direction of the
field changes faster than the redistribution process (i.e. as a result of an AC bias) the opposite effect does not show up.

In summary, we presented results of the time dependent conductivity
of insulating granular metals after excitation with a high electric
field. We find that a disordered granular system behaves differently
than its continuous counterpart. We interpret the results as
originating from a competition between two different trends. The
first process is the relaxation of the systems toward equilibrium
and is typical of all hopping systems. The second is a consequence
of charge rearrangement on the metallic grains and is expected only
in granular systems. In a sense, the granular metal excited by an
electric field is similar to spin glass subject to a sudden magnetic
field change. Despite the fact that a spin flip occurs on times of a
few ns, the magnetic interactions in the system cause the
arrangement of the whole system to occur on timescales which are
many orders of magnitude larger. We are grateful for helpful
discussions with O. Agam, V. Orlyanchik, Z. Ovadyahu, R. Berkovits
and M. Goldstein. This research was supported by the Israel Science
foundation (grant number 249/05).

\end{document}